# CHECKING COMPLEX NETWORKS INDICATORS IN SEARCH OF SINGULAR EPISODES OF THE PHOTOCHEMICAL SMOG.

Carmona-Cabezas Rafael[1,*], Gómez-Gómez Javier[1], Gutiérrez de Ravé Eduardo[1], Jiménez-Hornero Francisco J.[1]

[1] Complex Geometry, Patterns and Scaling in Natural and Human Phenomena (GEPENA) Research Group, University of Cordoba, Gregor Mendel Building (3rd floor), Campus Rabanales, 14071 Cordoba, Spain

* Corresponding author. e-mail: f12carcr@uco.es

ABSTRACT

A set of indicators derived from the analysis of complex networks have been introduced to identify singularities on a time series. To that end, the Visibility Graphs (VG) from three different signals related to photochemical smog ($O_3$, $NO$ concentration and temperature) have been computed. From the resulting complex network, the centrality parameters have been obtained and compared among them. Besides, they have been contrasted to two others that arise from a multifractal point of view, that have been widely used for singularity detection in many fields: the Hölder and singularity exponents (specially the first one of them).

The outcomes show that the complex network indicators give equivalent results to those already tested, even exhibiting some advantages such as the unambiguity and the more selective results. This suggest a favorable position as supplementary sources of information when detecting singularities in several environmental variables, such as pollutant concentration or temperature.

SP: Shortest Path; VG: Visibility Graph

1. INTRODUCTION

Photochemical smog is a severe problem that has gain attention of the scientific community in the last years. It is compounded by several gases and particles that have complex interactions. It becomes especially dangerous in highly populated and warm cities. One of the most recently studied gases is the tropospheric ozone due to its abundance, which makes it one of the main photochemical oxidants. It is a secondary pollutant, which in high concentrations, can affect human health and crops harshly (Doherty et al., 2009), as well as having a great impact on economy (Miao et al., 2017). It has been demonstrated that it does not only affect big cities, but also rural areas (Domínguez-López et al., 2014). Another interesting component of the photochemical smog is the nitrogen dioxide, which is a precursor for the mentioned ozone. It is a primary pollutant derived directly from the anthropogenic activity that arises in urban areas. It also has serious impacts on human health (Yue et al., 2018).

The World Health Organization (2005) stablished references for this kind of pollutants, in order to warn against dangerous effects. For instance, the level at which the ozone concentration is considered to be hazardous is 120 µg/m$^3$. For that reason, the identification of singularly high episodes of pollutant concentration gains importance.

Multifractal analysis has been previously used both for global behavior of the system and singularity detection in relation to pollutant dynamics (Pavón-Domínguez et al., 2015). In order to identify singularities in a general signal, one of the most commonly used techniques is the so-called pointwise Hölder exponents (Shang et al., 2006). It gives an estimation of how singular a given point is within a series, although its implementation has several disadvantages, as the dependence on parameters chosen by the user and numerical instability. Another one called the singularity exponent (Dai et al., 2014) will be used as well, in order to support the results from multifractal analysis.

In the last decade, a new approach designed to analyze time series was introduced by (Lacasa et al., 2008) and named Visibility Graph (VG). It is based on the transformation of those signals into a completely different mathematical object: a complex network. For the description of such new entities, the centrality parameters are very useful, as will be shown later in the text. Among the advantages of VGs, authors would like to remark the following: i) The characteristics of the original time series are inherited by the resulting network, leading to the possibility of describing the system through it. ii) They allow the analysis of various variables simultaneously, which can be used to find correlations.

In the introduced work, the VGs of three different time series related to the photochemical smog are computed. From them, the centrality measurements have been obtained and compared to multifractal indicators that are known to be useful for singularity detection. Finally, the searched purpose is to discern whether the complex network indicators can be used for the same applications that these multifractal parameters, giving equivalent results and overcoming some of their disadvantages.

## 2. MATERIALS AND METHODS

### 2.1. Data

This manuscript has employed real data from a 1 hourly ozone and nitrogen dioxide concentration (chemical factors from photochemical smog) and temperature (physical factor) time series, all measured in 2017. They were collected at the urban station in San Fernando (36°27' N, 6°12' W), in the province of Cádiz belonging to the southern part of the Iberian Peninsula. The reason behind choosing this location was that the area presents characteristics to be potentially vulnerable to the accumulation of photochemical smog (Domínguez-López et al., 2014). These are orographic (the Guadalquivir Valley), anthropic (two relevant industrial centers such as the chemical focus of Huelva and the Bay of Algeciras, and four capitals) and weather conditions (high solar radiation and temperature). The cited station is part of the network in charge of controlling the air quality in the region of Andalusia, which is administered by the Consejería de Medioambiente (Regional Environmental Department) and co-financed by the European Union.

To work with uncorrelated data (in order to obtain independent results), authors have selected different periods of time for each time series. For the case of ozone, the amount of data corresponds to the month of July; for $NO_2$, January has been chosen and, finally, for temperature data, October has been picked. Apart from the uncorrelation of the data, the choice was motivated by several reasons: in the case of ozone, the month of July presents the most suitable (and stable) conditions for the creation of this pollutant. For the nitrogen dioxide, January is the month where the photochemical reaction activity is lower and therefore its concentration depends more on the sources. Finally, October was chosen for the last series in order to see singular episodes of this quantity in a region where most of the year the oceanic influence stabilizes the temperature. This physical factor is unstable in autumn by nature in this area, as discussed in previous works (Dueñas et al., 2002). All real time series data have been represented in Figure 1.

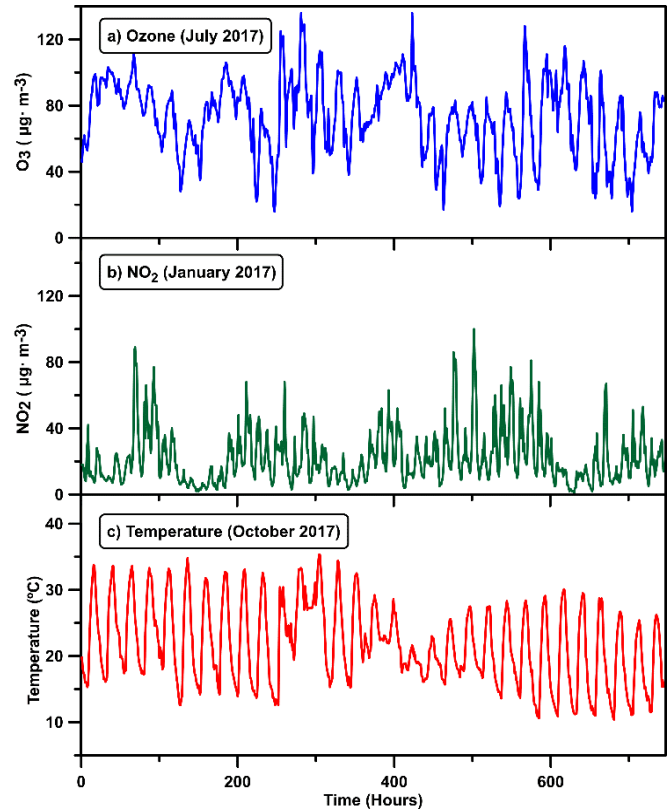

Figure 1: Ozone, nitrogen dioxide and temperature time series for each selected month.

## 2.2. Visibility Graphs

As mentioned before, the VG is introduced by (Lacasa et al., 2008) and is defined as a tool that makes possible to transform a time series into a graph, i.e. it converts a signal into a set of nodes connected through lines called edges.

To obtain the VG, which is associated to the time series, it is necessary to determine a criterion to establish which points (or nodes) are linked to each other, that is, have visibility. Let $(t_a, y_a)$ and $(t_b, y_b)$ be two arbitrary points from the time series which are chosen in order to check the mentioned criterion. One of the most commonly used is to consider that both have visibility (are connected in the graph) if any given point $(t_c, y_c)$ that is situated between the first two $(t_a < t_c < t_b)$ satisfies the following:

$$y_c < y_a + (y_b - y_a)\frac{t_c - t_a}{t_b - t_a} \tag{1}$$

This visibility algorithm concludes by repeating the previous step for every pair of points in the signal. As an example, one can observe this procedure applied to a sample time series in Figure 2.

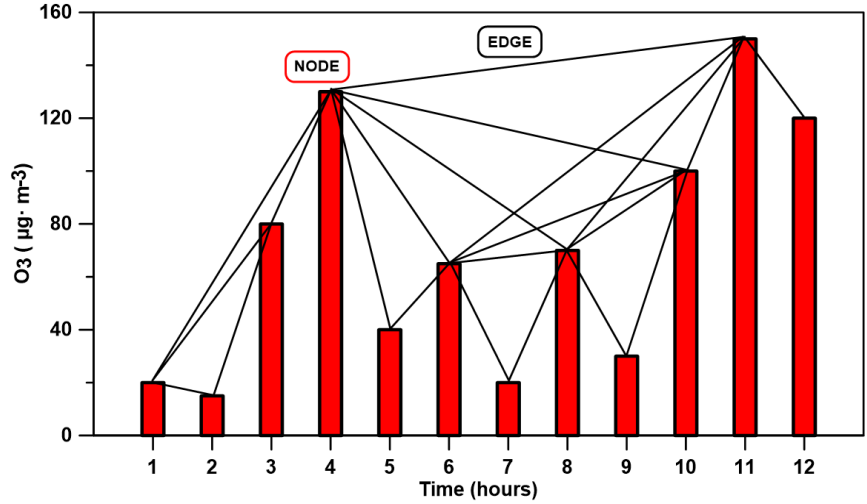

Figure 2: Visibility Graph obtained from a sample time series by the visibility algorithm. The nodes of the graph are the data points (red bars), while the links among them are illustrated as solid lines.

In practice, it is more useful to obtain a matrix representation of the graph that contains the information of the complex network: the visibility adjacency matrix. It is a $N \times N$ binary matrix, with $N$ the total number of nodes. Each element of the matrix $a_{ij}$ takes the value of unit if nodes $i$ and $j$ have visibility; otherwise, it is null and this means that nodes are not linked to each other.

The algorithm can be simplified after some factors are taken into account, leading to a visibility adjacency matrix with a general form as follows:

$$A = \begin{pmatrix} 0 & 1 & \dots & a_{1,N} \\ 1 & 0 & 1 & \vdots \\ \vdots & 1 & \ddots & 1 \\ a_{N,1} & \dots & 1 & 0 \end{pmatrix} \qquad (2)$$

Visibility Graphs are undirected networks, since the visibility criterion is reciprocal (one node has visibility with other and vice versa). However, there are some ways of mapping a directed network where the time order can be considered and this has been used previously for reversibility studies of time series (Lacasa et al., 2012; Xie et al., 2019). Nonetheless, when selecting a direction in order to account for the time order, there is a problem that arises regarding the mapped complex network. Since the size of the series is by definition finite, the (ingoing or outgoing) degree depends on the position of the point with regard to the end and beginning of the series. Ingoing degree refers to the number of links that enter into a node, while the opposite is for the outgoing degree. This means that the first points in the series will have more outgoing degree than the last ones, and vice versa. Due to this artifact, it not suitable for being used when a pointwise description (for singularity detection, for instance) is desired.

2.3. Complex networks indicators

Once the new complex network is retrieved, there are some parameters which can be studied to characterize its nodes importance, such as centrality measures. This concept was firstly used in the study of social networks to turn out to be introduced into other fields of knowledge (Agryzkov et al., 2019; Joyce et al., 2010; Liu et al., 2015). Some of these centrality measures that are used in this work will be further explained next.

2.3.1. Degree

The degree of a node ($k_i$) in a graph or complex network is defined as the number of other nodes with which it is linked ($k_i = \sum_j a_{ij}$). As authors exposed in the section 2.2, in the context of the visibility algorithm, two points having visibility means that they are linked because they fulfill the given criterion (see Equation 1). After computing the degree for each node, one can retrieve the probability or distribution for each result by using a histogram. This outcome is called degree distribution of the sample $P(k)$. There are some points with a singularly high degree, called *hubs*, that are of great importance in this distribution.

As it is shown in previous works, the degree distribution obtained from the VG can characterize the nature of the time series involved (Lacasa et al., 2008; Mali et al., 2018). For example, it is possible to make a distinction among fractal, random or periodic signals.

### 2.3.2. Closeness

In the previous sections, centrality parameters have been defined by considering the number of edges and the adjacency matrix properties. Nevertheless, it is necessary to specify the meaning of another property within a graph in order to carry on with next definitions, which is the so-called shortest path (SP). In a network, one can observe a different number of edges (as a measurement of length) passing through any (in general, distant) pair of nodes. Two distant nodes $(i, j)$ will have different number of edges and paths between them, but there will be some of these paths where the number of edges will be minimum; this quantity is named as the SP.

If one takes all pairs of nodes, it is possible to obtain a matrix, the so-called distance matrix $D$, where each element $d_{i,j}$ contains the SP from node $i$ to $j$. One usually sets diagonal elements as zero. For an undirected graph, this matrix will be symmetric, as in the adjacency matrix case (see Section 2.2).

After the explanation of this graph property, it is possible to define the closeness centrality of a node $i$ as the inverse of the sum of distances from this node to the others:

$$c_i = \frac{1}{\sum_{j=1}^{N} d_{i,j}} \qquad (3)$$

Where $d_{i,j}$ is the element $(i,j)$ from the corresponding distance matrix of the graph.

### 2.3.3. Betweenness

The main idea behind betweenness parameter is to focus on the centrality as a measurement of how a node is between many others. That is, how much a node is passed through by shortest paths of other pairs of nodes. Therefore, the equation that defines this quantity for a node $i$ is the following:

$$b_i = \sum_{\substack{j=1 \\ j \neq i}}^{N} \sum_{\substack{k=1 \\ k \neq i,j}}^{N} \frac{n_{jk}(i)}{n_{jk}} \qquad (4)$$

Where $n_{jk}$ is the number of SP's from $j$ to $k$ (notice that these paths can be degenerated), whereas $n_{jk}(i)$ is the number of SP's that contains the node $i$.

### 2.4. Multifractal indicators

### 2.4.1. Pointwise Hölder exponent method

The Hölder exponent is a measure used in the context of multifractal analysis to characterize the singularities which are present in a signal. A function or a time series is fractal when it exhibits some local properties as self-similarity, irregularity, fine structure and fractional dimension; if they are variable at different points, then this function is multifractal. Consequently, multifractal analysis describes the singularities implicated in a time series.

A commonly used method for multifractal analysis of a signal is to compute the pointwise Hölder exponent. This parameter is defined as a local characteristic of a function which is computed at every point of its domain. It refers to the decay rate of the amplitude of the function fluctuations in the neighborhood of the data point when the size of the neighborhood shrinks to zero, that is, the function singularities. The Hölder exponent at a point $t$ of $f(t)$ can be expressed as:

$$\alpha_f(t) = \lim_{h \to 0} \inf \frac{\log|f(t+h) - f(t)|}{\log|h|} \qquad (5)$$

In practice, one can only obtain discrete time series and so some different methods for computing the Hölder exponent has been elaborated to solve this problem. (Peng-Jian and Jin-Sheng, 2007; Shang et al., 2006) developed an algorithm for numerical evaluation of Hölder exponent based on the previous equation. This method takes $n+1$ points from a signal equally spaced, $\{y_0, y_1, \dots, y_n\}$ and calculates the intensity of its Hölder exponent at a specific point $y_i$. This computation considers a number of preceding and following points, $s$, which is named as the window width and is set by the user (in total, $2s$ values around the point are taken). Each value in the window has got a different weight controlled by other parameter, $\lambda$, known as regression coefficient and whose values are in the interval $[0, 1]$. This last parameter reduces the weight for each point in the window as they are further from the center. Their influence over the ultimate computation of the Hölder exponent must be greater for close points than those which are more distant.

Once values for $\lambda$ and $s$ are selected, it is necessary to obtain the next quantity for each integer $k \neq 0$ from $-s$ to $s$ in each window:

$$R_{i,k} = \frac{\log\left(\frac{y_{i+k} - y_i}{C_1}\right)}{\log\left(\frac{|k|}{C}\right)} \qquad (6)$$

Where $C_1$ and $C$ are the normalizing parameters which are chosen for the convenience of computation (Peng-Jian and Jin-Sheng, 2007) and $k$ must fulfill $0 \leq i + k \leq n$. For points which are at the very beginning or end of the signal, where the $k$ index falls out of the domain, the window width $s$ is properly shrunk. Next, for each integer $j$, $0 \leq j \leq s$, it must be computed:

$$h_{i,j} = \min\{R_{i,k} : |k| \leq j\} \qquad (7)$$

This last equation is related to take the lim inf as in the Equation 5 for the case of a continuous function. Finally, the local Hölder exponent, $\alpha_i$, is retrieved by the computation of the weighted average of the approximations $h_{i,j}$:

$$\alpha_i = \frac{1 - \lambda}{1 - \lambda^s}\left(h_{i,1} + \lambda h_{i,2} + \lambda\ h_{i,3} + \cdots \right. \\ \left. + \lambda^{s-1} h_{i,s}\right) \qquad (8)$$

Where the most important factors are those which are close enough to ith point.

### 2.4.2. The singularity exponent method

In an effort to get better results for the singularities of traffic data from a highway, (Dai et al., 2014) proposed another approach to analyze this property through the "singularity exponents", as they named in the paper.

The algorithm takes a given time series data of length $n$, $\{y_1, y, ..., y_n\}$, uniformly spaced, and requires choosing two window widths $r_i$ ($i = 1, 2$). Next, one must compute the following quantities:

$$\bar{y} = \frac{1}{n} \sum_{k=1}^{n} y_k \tag{9}$$

$$\overline{y_{r_i}(t)} = \frac{1}{2r_i + 1} \sum_{k=-r_i}^{r_i} y_{t+k} \tag{10}$$

$$|\Delta y(t)|_{r_i} = \left| \overline{y_{r_i}(t)} - \bar{y} \right| \tag{11}$$

Where $\bar{y}$ is the average of the whole signal, whereas $\overline{y_{r_i}(t)}$ and $|\Delta y(t)|_{r_i}$ denotes the average volume in the field and the absolute errors between $\bar{y}$ and $\overline{y_{r_i}(t)}$ in $[-r_i, r_i]$, respectively. It must be noticed that Equation 10 must satisfy $0 \leq (t + k) \leq n$.

Finally, the singularity exponents at each point $\beta(t)$ are obtained by computing the fluctuations in each of the previously defined scales:

$$\epsilon_{r_1}(t) = \frac{1}{2r_1} \sum_{t'=t-r_1}^{t'=t+r_1} |\Delta y(t')|_{r_1} = c(t) r_1^{\beta(t)} \tag{12}$$

$$\epsilon_{r_2}(t) = \frac{1}{2r} \sum_{t'=t-r_2}^{t'=t+r_2} |\Delta y(t')|_{r_2} = c'(t) r^{\beta(t)} \tag{13}$$

Assuming that $c(t) \approx c'(t)$, then, one can get:

$$\beta(t) \approx \frac{\ln \epsilon_{r_1}(t) - \ln \epsilon_{r_2}(t)}{\ln r_1 - \ln r} \tag{14}$$

Finally, a usual way of checking the Hölder exponent method is to take the generalized Weierstrass function as a test function. In this work, authors have proven the implemented algorithm of both multifractal indicators. The selected Weierstrass function is the following:

$$f(t) = \sum_{k=0}^{\infty} 3^{-ks(t)} \sin(3^k t) \tag{15}$$

Where $s(t)$ is the seed function, whose values are contained in the interval $[0, 1]$. As (Daoudi et al., 1998) shows, $s(t) = \alpha_f(t)$ for all $t$ and so, one can compare theorical values of the Hölder exponent (given by $s(t)$) with the numerical results.

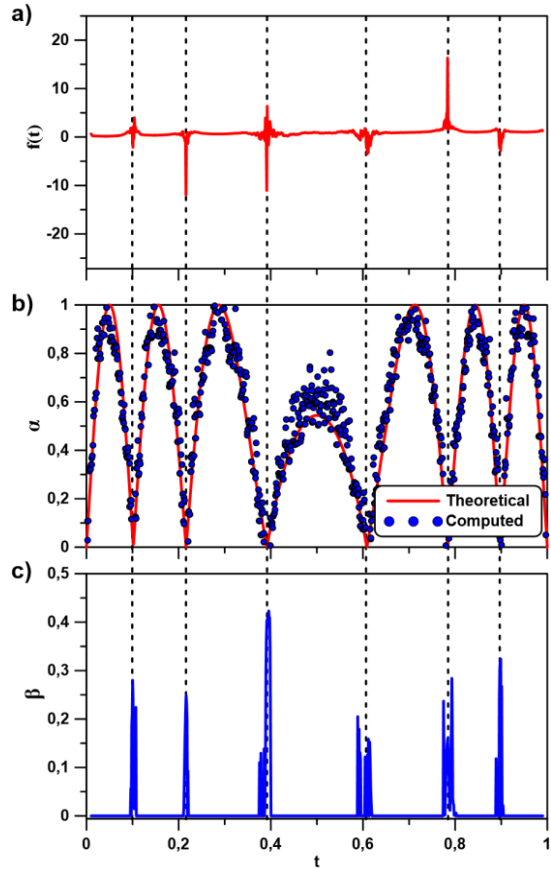

Figure 3: Time series obtained from the Weierstrass function for an interval $t \in [0, 1]$ (a) and the Hölder and singularity exponents $\alpha$ and $\beta$ (b and c, respectively).

This analysis can be observed in the Figure 3 for both algorithms, where the chosen seed function is $s(t) = |\sin(10\sin(\pi x))|$. As expected, the Hölder exponent method fits well the theorical value, while the singularity exponent method does it as well. However, in the last case, shown in Figure 3c), the level of singularity of the data can be understood as the deviation from 0, which is independent of the sign. For this reason, authors have decided to plot the absolute value of the results.

## 3. RESULTS AND DISCUSSION

In this section, the parameters previously introduced in the last part are tested on different real time series of environmental variables related to photochemical smog (see Section 2.1). As it was mentioned before, these three time series have been chosen based on the fact that they are different in nature, in order to test the proposed indicators for distinct scenarios. All the complex network indicators shown in the plots are normalized to the maximum value of each one of them.

In order to evaluate these quantities, the widely used Hölder exponents have been obtained following the approach described in the former section. The parameters set for the calculation have been in all the cases $\lambda = 0.8$, $C_1 = 70$, $C_2 = 10^5$ and $s = 10$. Moreover, the other exponent proposed in order to overcome some of the shortcomings of the Hölder exponents (Dai et al., 2014) has been analyzed (the singularity exponent). For this indicator, the chosen parameters were $r_1 = 1$ and $r_2 = 7$.

In Figure 4, all these indicators are shown for the case of the ozone concentration time series. For the other series, the results are similar and therefore the same considerations are taken into account. It must be pointed out that the complex network ones are univocal for a given time series, since the VG constructed does not depend on any numerical parameter. On the other hand, the multifractal indicators depend substantially on the chosen parameters when running the algorithm (see Sections 2.4.1 and 2.4.2).

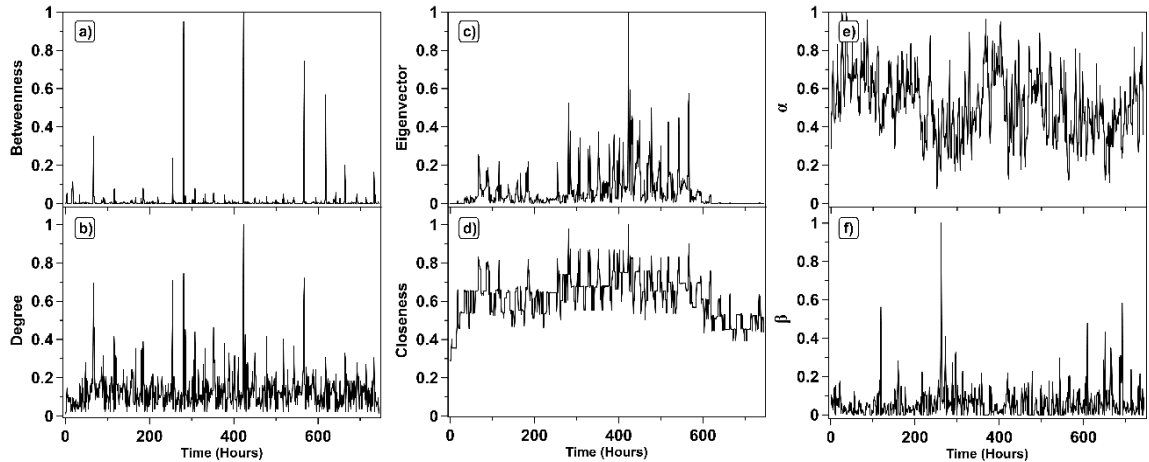

Figure 4: Different indicators computed from the whole ozone concentration time series: (a – d) Centrality parameters from the Visibility Graph and (e - f) Hölder and singularity exponents.

In order to make Figure 4 more intelligible, the followed process is explained next. What has been done in practice is computing firstly the betweenness centrality of the total time series. Afterwards, a dynamic criterion has been stablished in order to select the most important central nodes from this quantity. It consisted on searching the relative maxima which are above a given percentage of the absolute one. After several tests, it has been found that the identification of singularities holds down to a 5%, which could be used as a rule of thumb for future works. Nonetheless, with this value, a considerable amount of peaks are chosen and for practical reasons, from this point only the five most central nodes will be shown in the figures. These points correspond to the five main *skyline hubs* (Carmona-Cabezas et al., 2019b) in the series (i.e. the nodes with the highest singular values of betweenness). The reason for choosing this term was the similarity to the skyline drawn by the skyscrapers in a city from the point of view of other nodes.

Once those nodes are selected, authors have investigated the values of the rest of the indicators around them. The reason for choosing the betweenness centrality as a first reference is the more distinguished and smoother results that it provides for some specific important nodes from the VG, as depicted in Figure 4a). Furthermore, the multifractal measurements are built upon parameters chosen by the user for the convenience of the computation. This might lead to problems and ambiguities in the outcomes.

Firstly, the results obtained for the ozone concentration time series can be regarded in Figure 5, where five different betweenness peaks (*skyline hubs*) from b) were closely studied (numbers 1 - 5). The different complex network indicators have positive pronounced peaks in the same temporal points of the series. When it comes to the Hölder and singularity exponents from multifractal analysis, they present minima and maxima values respectively at those same points as well.

In order to understand the relation of this parameter to the photochemical pollution (ozone in this case), authors would like to point out a recent study (Carmona-Cabezas et al., 2019b). In that work, betweenness centrality peaks have been related with the values of the series that store most of the information about the upper envelope. Considering this, a skyline hub can be regarded as a singular episode of ozone concentration in the sense that it indicates a change of the tendency with respect to the previous maxima. For instance, if the maximal concentration of ozone of several days is steadily increasing and then starts to decrease, that critical point is identified by a betweenness peak.

In Figure 5c1), the peaks selected by the authors for their higher betweenness are magnified and superposed in the same plot. This has been done in order to compare with the rest of parameters. As it could be seen as well in Figure 4, betweenness centrality gives the clearest signal and therefore its peaks are sharper.

The next indicator (see Figure 5c2) from the VG is the degree of the nodes, which has been widely used in many studies of this kind (Carmona-Cabezas et al., 2019a; Pierini et al., 2012; Zhou et al., 2017). Before studying its usefulness for ozone description, it is easily regarded at a glance that all the positions selected correspond as well to very pronounced peaks in the degree (*hubs*). It is known that a high degree implies a high concentration of ozone (Carmona-Cabezas et al., 2019b), while the opposite is not always true. Therefore, when a point is identified as a singularity from the degree, it means that the ozone concentration at that particular time is especially high in magnitude within a time interval around it. It suggests that the conditions for its production would be optimal at that time. The correspondence with the betweenness peaks is due to the fact that when ozone concentration reaches a singular maximum before a downwards trend, this peak magnitude will be as well singularly high, locally.

The last of the VG indicators is the closeness centrality. This quantity has been used before to describe VGs with interesting results, although in theoretical point of view (Bianchi et al., 2017; Iacovacci and Lacasa, 2019). Closeness centrality of a given time point is related to the values at its left and right (Donner and Donges, 2012). Thus, authors attribute a singular peak in the closeness to a rarely high episode surrounded by a concave up tendency (since it favors connectivity) in the surrounding concentrations of this pollutant. As depicted in Figure 5c3), the betweenness peaks coincide as well with singular high values of closeness, in every one of the selected points. Nevertheless, the obtained curves are rather noisy (compared to the degree and betweenness), although the peaks can be inferred without problem.

Finally, the multifractal indicators are shown in Figure 5c4) and c5). On the one hand, the Hölder exponent plot displays minimal values around the positions used as reference. It was expected, since highly irregular points exhibit Hölder exponents closer to zero, whereas smoother regions present greater values (Jaffard, 1997; Safonov et al., 2002). On the other hand, the singularity exponents, as explained before in Section 2.4.2, show maximal values on singular points. That is as well corroborated in the reference peaks positions, in a clearer way than the Hölder exponents in this case. In any case, neither of the multifractal indicators show peaks as acute as the complex network ones. It should be pointed out as well the noise embedded within the singularity exponent curves, that makes more difficult to identify the searched maxima for some of the peaks (for instance, peak 1 and 2).

When it comes to the underlying tropospheric ozone pollution concentration, all of these singularities correspond to unusual episodes of especially steep accumulation of that gas. More precisely, the selected peaks coincide with $3^{rd}$, $12^{th}$, $18^{th}$, $24^{th}$ and $26^{th}$ of July, all of them located between 2 PM and 6 PM (GMT+2), which is the time period when the radiation and temperature reach their maxima. I every case, the corresponding concentrations are above or reaching the dangerous threshold stablished by the World Health Organization. What is significant about the selected ones is that, as can be regarded in Figure 5a), they are placed before a change in the tendency of the previous maxima or daily concentrations of ozone. It corresponds to what was previously discussed for betweenness, as the peaks are selected by looking at this quantity. The utility of this is that once the highly irregular $O$ concentration that diverge abruptly from the trend are detected, it would be possible to perform a deeper study in order to find the origin of it and act accordingly to prevent future similar episodes.

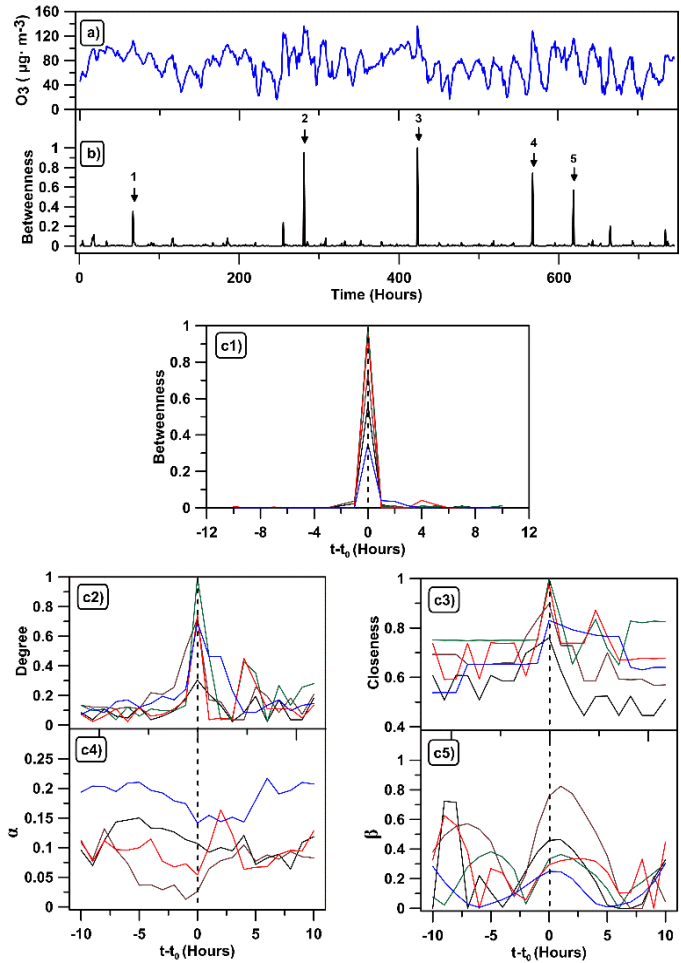

Figure 5: Ozone concentration time series (a) with the betweenness values computed from it below (b). Plots from c1) to c3) show the complex network indicators: betweenness, degree, closeness (in appearance order). c4) and c5) display the Hölder and singularity exponents (respectively).

In Figure 6, the same parameters as in the previous case are studied for the $NO$ concentration time series. The same procedure was followed in order to obtain the plots. It can be regarded that equivalent behaviors are present here: all peaks from the complex network indicators identify quite well the singularities found in betweenness; whereas the multifractal ones (which coincide as well) have wider shapes, with less accuracy. Nonetheless, the Hölder exponent minima (see Figure 6c4) are much clearer than above. The curves corresponding to the singularity exponent exhibit again a considerable level of noise.

As in the previous case, looking at the physical meaning of the series, the peaks where authors focus accord with unusual maxima of $NO_2$ throughout the month. In detail, the days corresponding to these pollutant concentration singularities are 3$^{rd}$, 11$^{th}$, 17$^{th}$, 21$^{st}$ and 28$^{th}$ of January. In this case, the singular hours are not as consistent as in the ozone. Authors attribute this irregularity to the fact that the main source for $NO_2$ is human activity, which in many cases might vary. Hence this could be used to identify singular acute activity from industry or traffic, as well as meteorological unexpected events, that could lead to unanticipated concentrations derived from transport of this pollutant.

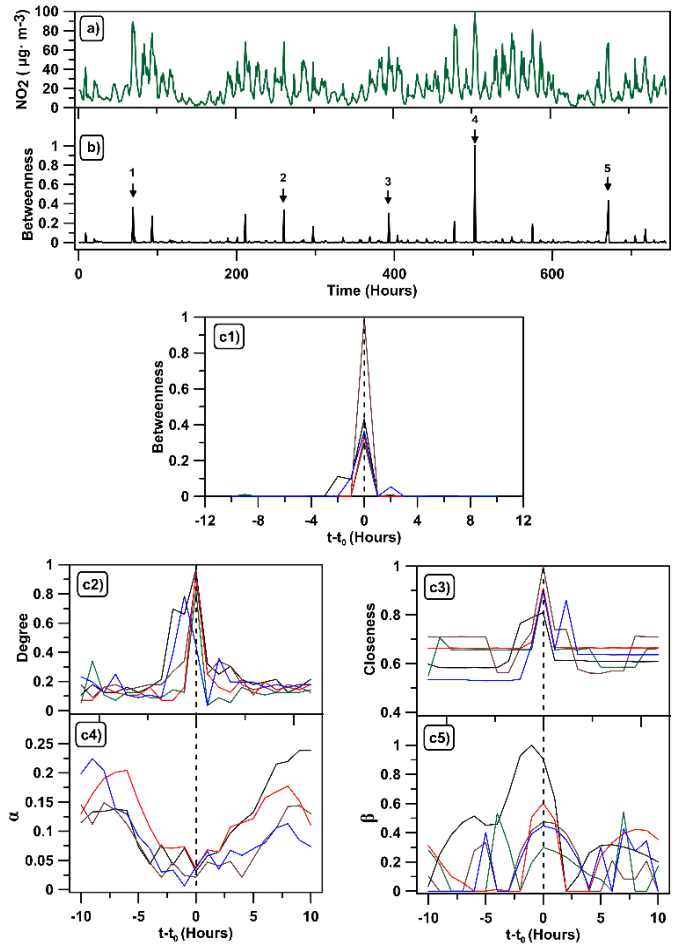

Figure 6: $NO_2$ concentration time series (a) with the betweenness values computed from it below (b). Plots from c1) to c3) show the complex network indicators: betweenness, degree, closeness (in appearance order). c4) and c5) display the Hölder and singularity exponents (respectively).

Finally, a different type of time series has been analyzed through all the indicators shown up to this point. In this case, this series corresponds to the hourly average temperature measured, which as expected, displays a more regular behavior (see Figure 7a). The complex networks parameters identify in a similar way the same singular points. Moving to the multifractal parameters, the singularity exponent behaves as in the previous series (see Figure 7c5), in contrast to the Hölder exponent, that shows maxima instead of the expected minima (see Figure 7c4). Authors attribute this anomaly to some of the disadvantages of the computation of this parameter by the algorithm, which under some circumstances may provide non-finite or misleading values.

As mentioned for the previous variables, the actual meaning of these peaks resides on unexpected values of high temperature that occur on the 6$^{th}$, 13$^{th}$, 26$^{th}$, 27$^{th}$ and 28$^{th}$ of October. Now, the singularities encountered on temperature might be associated to unpredicted meteorological events.

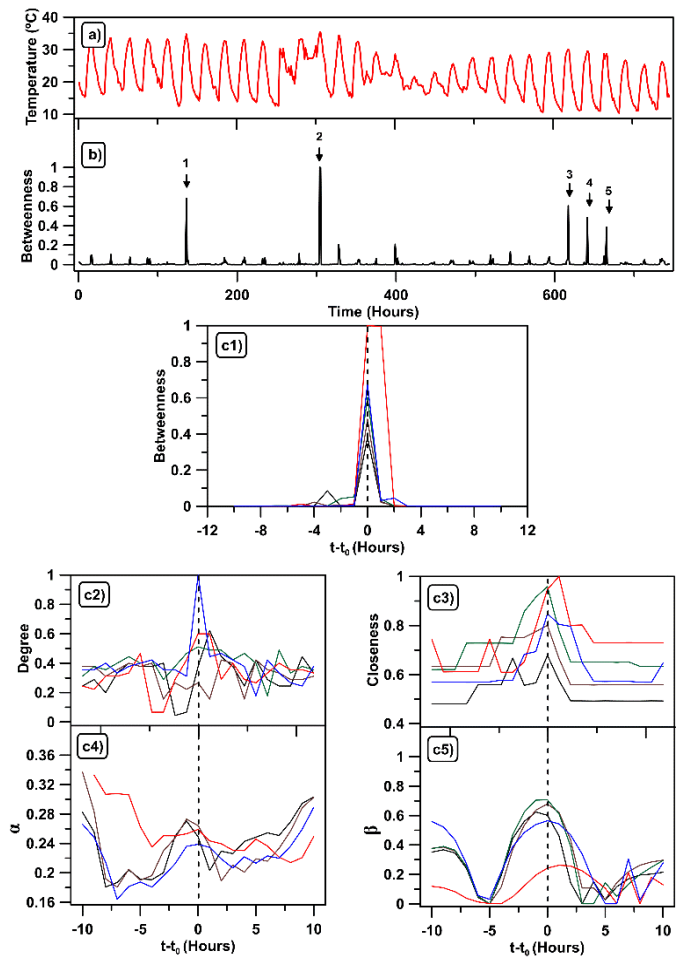

Figure 7: Temperature time series (a) with the betweenness values computed from it below (b). Plots from c1) to c3) show the complex network indicators: betweenness, degree, closeness (in appearance order). c4) and c5) display the Hölder and singularity exponents (respectively).

## 4. CONCLUSION

After discussing the proposed indicators, authors consider that these complex network parameters can be properly used to identify relevant points in environmental time series, such as the ones analyzed here. It has been demonstrated how indicators that are different in nature, can obtain complementary results that can be employed to characterize the behavior of experimental signals from pollutants and temperature in the context of photochemical smog. They have been compared to widely known singularity indicators from multifractal analysis, showing some advantages as well. This opens a bridge between complex networks and multifractal studies for local singular behavior of time series.

Finally, it can be argued that some of the shortcomings of the Hölder and singularity exponents are overcome with the proposed methodology. On the one hand, the multifractal indicators depend on a number of parameters that rely on the nature of the series, giving different ambiguous results. By contrast, complex network ones are univocal for a time series. No matter how one runs the algorithm, the result would be the same. On the other hand, the way Hölder exponents are defined gives complications for certain cases derived from the logarithm in its expression. Conversely, the centrality parameters do not have such problems, since their computation is based on simple arithmetic calculations from graph theory. Also, based on the properties of each parameter, they seem to be able to describe different properties of the pollutant dynamics at the selected times. For instance, betweenness is found to be related to a change in the tendency of the upper envelope of the signal; degree identifies singularly high concentrations or temperature episodes; while closeness characterizes the behavior of the concentrations surrounding the detected singularity. Taking all these facts into account, it is possible to consider the proposed indicators as a future additional information source.

For future works, it remains open a wide range of possible applications for these local studies. For instance, in the field of environmental analysis, authors would like to stress the possibility to employ these indicators to relate singular events of different variables, such as the ozone and some of its precursors (both chemical and physical). Also, based on the use of the Hölder exponent for predictive purposes (Shang et al., 2006), it might be investigated the usefulness of VGs for the same aim.

5. ACKNOWLEDGEMENTS

The FLAE approach for the sequence of authors is applied in this work. Authors gratefully acknowledge the support of the Andalusian Research Plan Group TEP - 957 and the XXIII research program (2018) of the University of Cordoba.

**DECLARATION OF INTERESTS**

Declarations of interest: none.

**HIGHLIGHTS**

- Visibility Graphs can be used to identify singularities in pollutant series.

- Peaks from complex network indicators coincide with the multifractal ones.

- Hölder and singularity exponents give ambiguous results due to parameter selection.

- Betweenness and degree give the clearest signal for identifying singular points.

- Among VG indicators, eigenvector centrality gives the least accordance.

**Graphical Abstract**

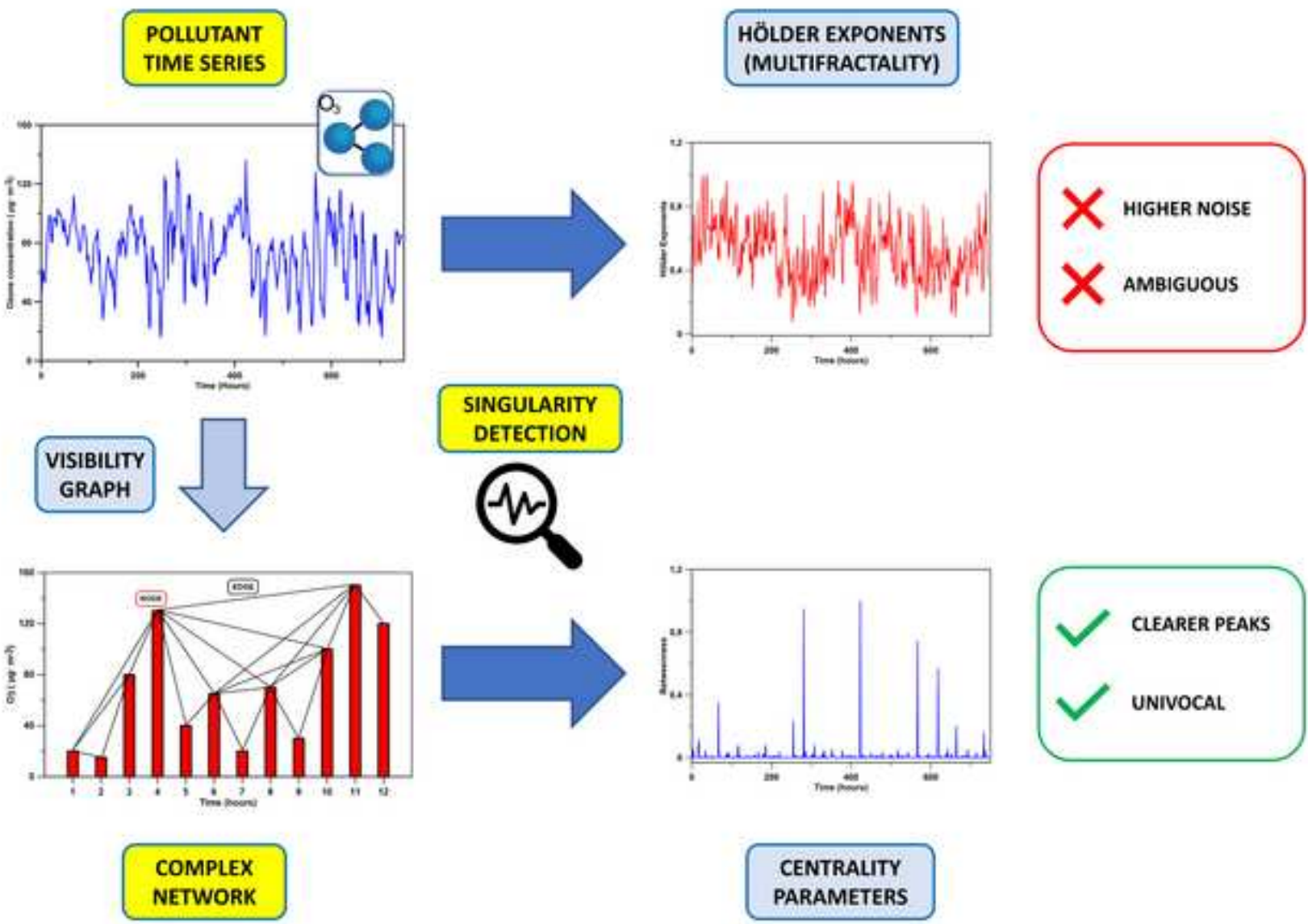